# Band structures and direct-to-indirect bandgap transitions in BAlN and BGaN alloys: a first principle study

**Che-Hao Liao, Feras AlQatari and Xiaohang Li***

*Advanced Semiconductor Laboratory, King Abdullah University of Science and Technology (KAUST), Thuwal 23955-6900, Saudi Arabia*

*Corresponding author: xiaohang.li@kaust.edu.sa

## Abstract

In this work, the energy band structures of BGaN and BAlN alloys are systematically studied through first-principles calculation using HSE hybrid density functional theory by MedeA-VASP. Direct-indirect bandgap transition of BGaN alloys at B content around 44% and that of BAlN alloys at B content about 24% have been identified. The variation of electron and hole effective masses of both materials at different B compositions have also been demonstrated. A large change in hole effective masses of BGaN and BAlN alloys from B=0% to 25% has been observed. Finally, a picture of energy bandgap versus lattice constant of III-nitride family with boron is shown.

## Introduction

Wide bandgap III-nitride materials have attracted great attention due to their excellent properties for device applications. Recently, boron-containing III-nitrides have been added to the family due to their potential in adding more tunability to the bandgap and lattice parameter, in addition to other electronic properties[1-9]. Moreover, it has been experimentally shown that a small incorporation of boron into GaN and AlN can cause a significant change in the refractive index of materials[10,11]. This allows them to be matched to different substrates such as AlN, SiC and $Ga_2O_3$. Typically, the ternary alloys BAlN and BGaN are studied, but more degrees of freedom are added when considering the quaternary BAlGaN alloys. Adding boron could open up many possibilities in designing sophisticated and reliable high-frequency, high-power micro- and optoelectronic devices such as high electron mobility transistors (HEMTs) and ultraviolet emitters. Unfortunately, the basic electronic and structural properties of these alloys are still not well known.

For high electron mobility transistors, BGaN layers could be used as semi-insulating buffer layer or back-barrier layer to provide a polarization-induced band discontinuity and a resistive barrier originating from its excellent insulation properties[12,13]. The electrical resistivity increases strongly by increasing the boron content in BGaN and the n-type carrier concentration decreases while the mobility increases[12]. This boron controlled resistivity is very promising for using the BGaN based materials in microelectronic and optoelectronic devices.

The growth of BAlN and BGaN is still challenging due to a large difference in atomic radius between boron and aluminum and gallium, which is expected to result in a wide miscibility gap and phase separation at several percent of boron. Growing BGaN layers by using different substrates have been demonstrated, and BGaN layers with the boron content up to 5.5% and 4.3% have been reported in BGaN deposited on SiC substrates and AlN/sapphire templates, respectively[14]. The highest reported boron content of 7% in BGaN film was obtained using ion implantation into GaN film grown on sapphire[15]. There is still room for crystal growth improvement and innovation in this field by exploring boron nitride-based alloys.

In this work, the band structures of $B_xGa_{1-x}N$ and $B_xAl_{1-x}N$ are systematically studied through first-principles calculation based on density functional theory (DFT) using hybrid functional of Heyd, Scuseria, and Ernzerhof (HSE) in the MedeA-VASP code. The band structures of the studied alloys with varying B compositions were calculated and poltted. The direct and indirect bandgap energies are extracted and plotted as a function of B composition, and a direct-indirect bandgap crossover is observed. Furthermore, the variation of direct and indirect bandgap energies are further broken down for the comparison at different high symmetry points. The direct-indirect bandgap transition of $B_xGa_{1-x}N$ alloys at B concentration around 44%, and at 24% for $B_xAl_{1-x}N$. Additionally, we extracted the electron and hole effective masses of both kinds of materials from their bandgap structures. The variations of effective masses at different B compositions have been demonstrated, and a drop in hole effective masses of BGaN and BAlN alloys from B=0% to 25% has been observed. Finally, a figure of Energy bandgap versus Lattice constant of Boron with III-Nitride family has been drawn through our first-principles calculation.

## Computational Method

In this work, the band structures of $B_xAl_{1-x}N$ and $B_xGa_{1-x}N$ alloys are systematically studied through first-principles calculation based on DFT. Since GaN is a direct-bandgap semiconductor and BN is indirect-bandgap, a crossover from direct to indirect bandgap will occur. The DFT calculations use hybrid functional of Heyd, Scuseria, and Ernzerhof (HSE) as implemented in the MedeA-VASP code. To cover different compositions of B, the supercell approach is applied. For the structure optimization process, all the structures have been fully relaxed using general gradient approximation (GGA-PBEsol) exchange-correlation with Hellman-Feynman force less than 0.02eV/Å [16]. The Brilluin Zone has the grid being gamma-centered with k-point meshes of 6×6×6 for both the 16-atoms supercells and for the primitive unit cells. The energy cutoff was set to 520 eV for the plane-wave basis set in all calculations.

## Results

The calculated band structures for the $B_xGa_{1-x}N$ system are shown below (Figure 1). Using linear interpolation between the calculated data points, we find that the direct-to-indirect bandgap crossover occurs around boron composition of 44% (Figure 2(a)). Figure 2(b) shows the bandgaps along different symmetry lines. These lines are between the Γ point for the valence band and the Γ, A and the K points for the conduction band. We can see that above 75% boron composition the slope of change of bandgaps along different symmetry lines increases.

To calculate the effective masses of electrons and holes near conduction bands minima and valence bands maxima, respectively, the band structures were calculated around those points with higher sampling. After that, the equation below was used to calculate those effective masses

(Figure 3).

$$E = \frac{1}{2} m v^2 = \frac{1}{2} \frac{p^2}{m} = \frac{(\hbar k)^2}{2m} \rightarrow m^* = \frac{\hbar^2}{\frac{d^2 E}{dk^2}}$$

The calculated effective light hole (lh) and heavy hole (hh) masses drop significantly between 0 and 25% boron composition (Figure 4). At higher compositions the effective masses increase at a much slower slope.

The same process is repeated for BAlN and the results are similarly show in Figures 3-6. The crossover point for BAlN was found to be near boron composition of 24%. The hh effective mass is found to also drop significantly at low compositions similar to what is found in BGaN. Similar to BGaN, the bandgaps at high boron compositions of BAlN are found to increase in slope.

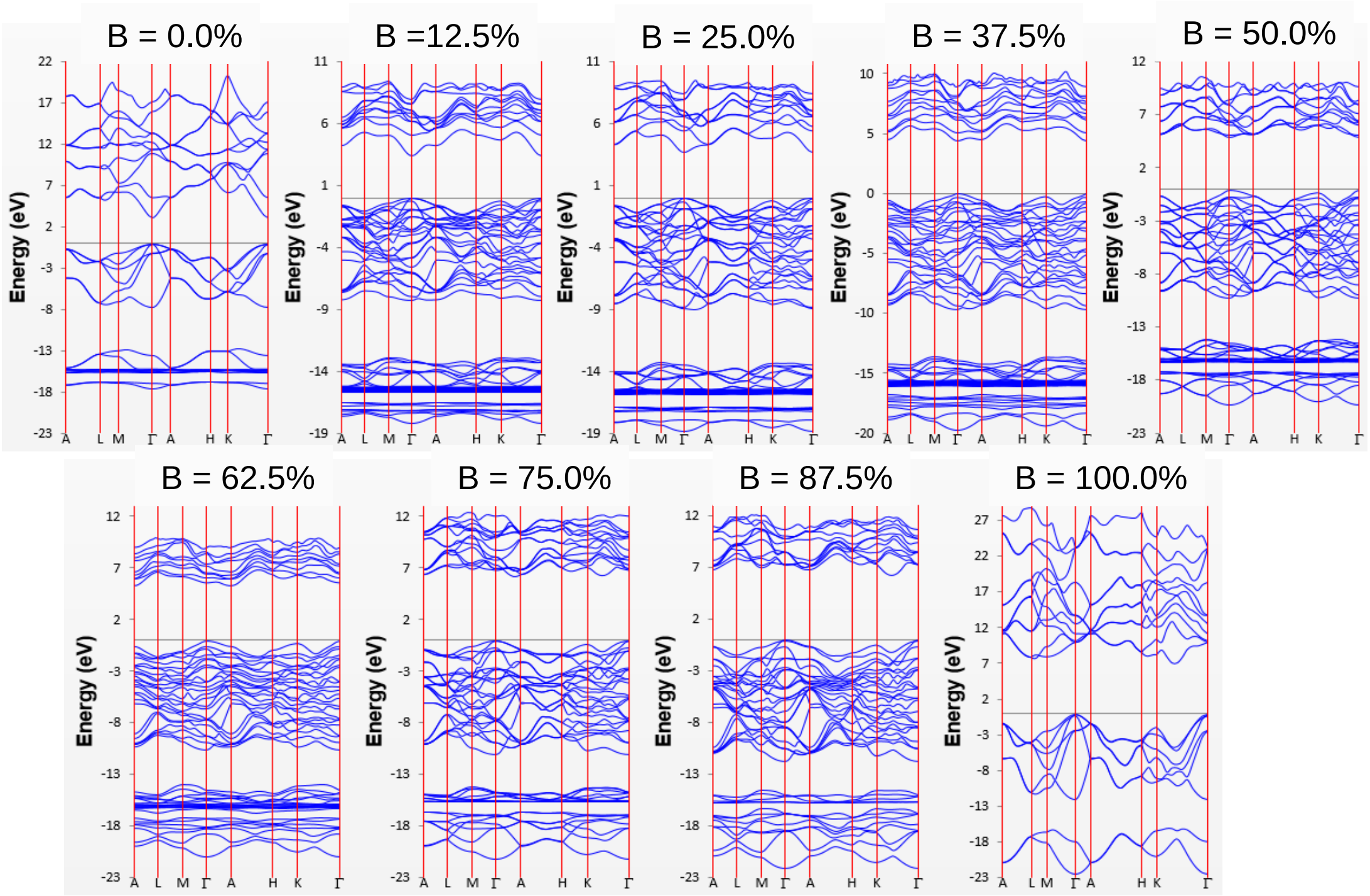

Figure 1 Electronic band structures of BGaN with different boron compositions.

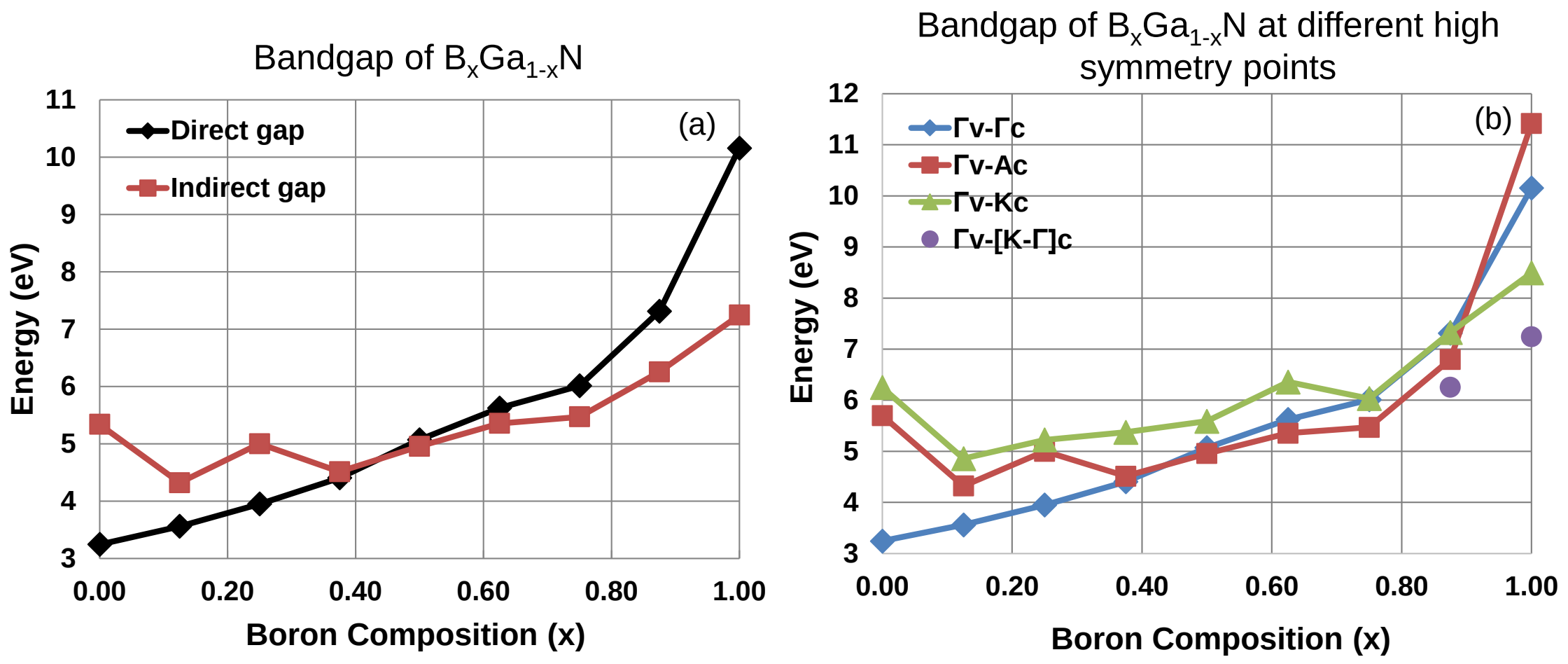

Figure 2 (a) The direct and indirect bandgaps of BGaN as functions of boron composition. (b) The bandgaps of BGaN for different transitions between chosen symmetry points.

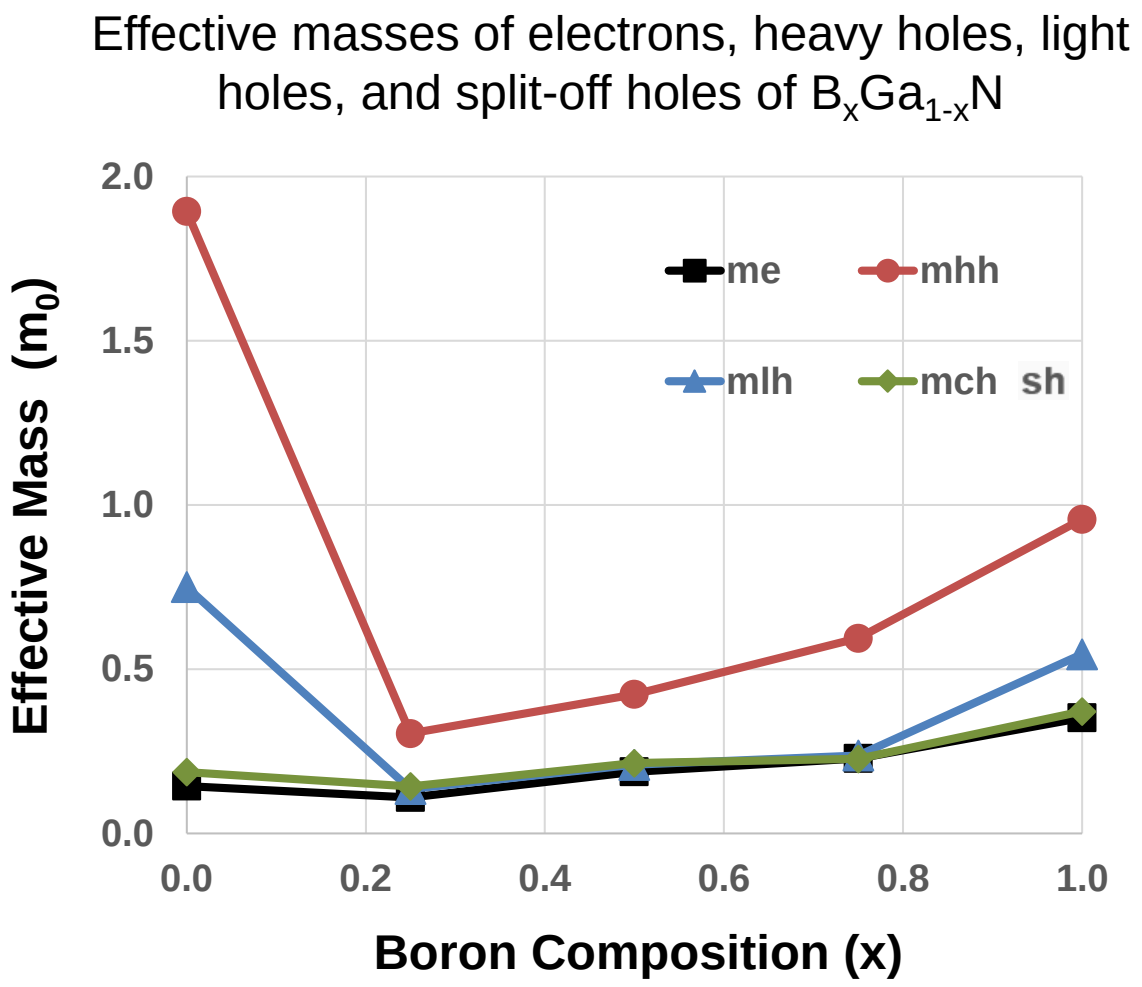

Figure 3 Effective masses of BGaN with different boron compositions.

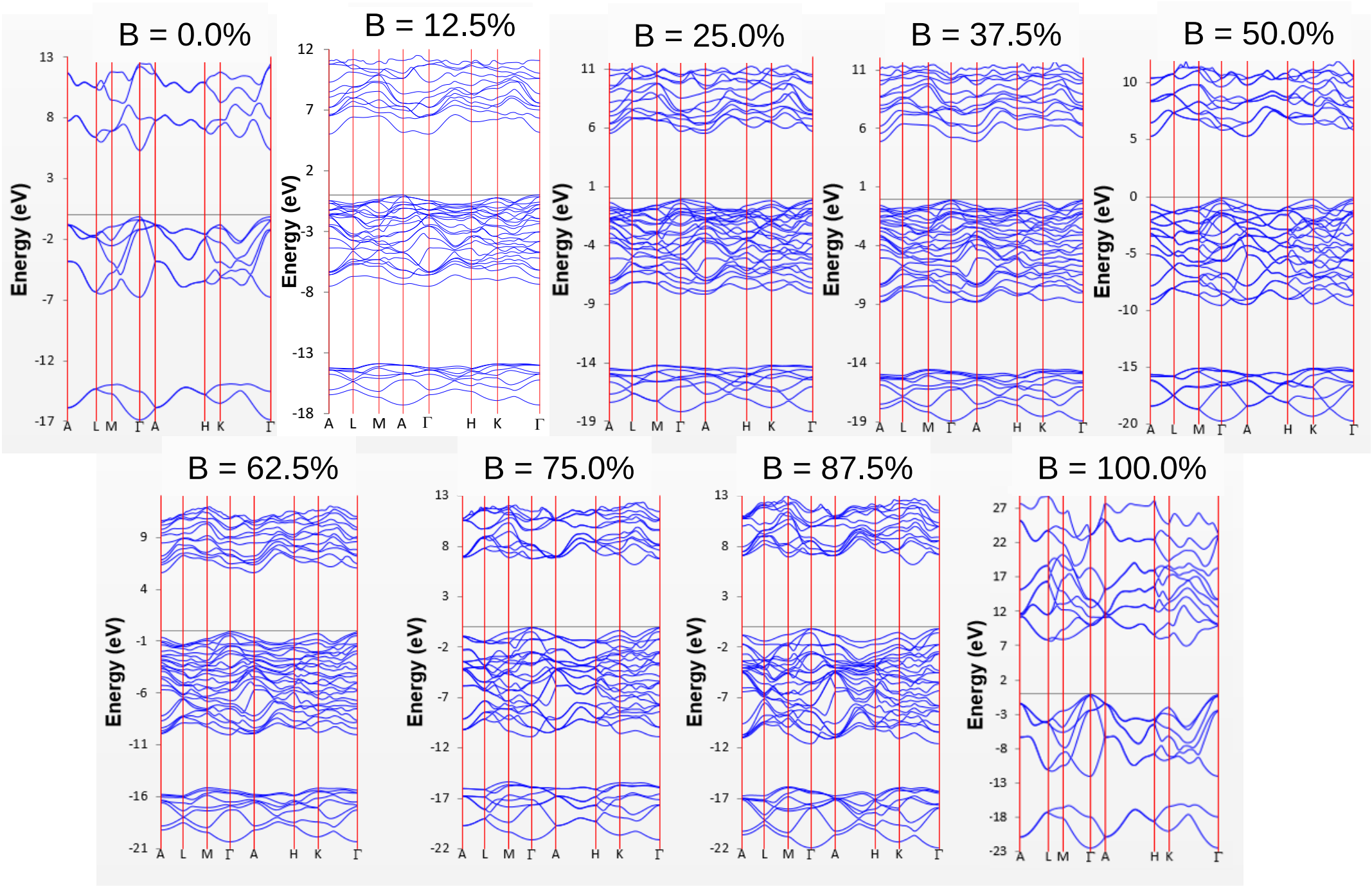

Figure 4 Electronic band structures of BAlN with different boron compositions.

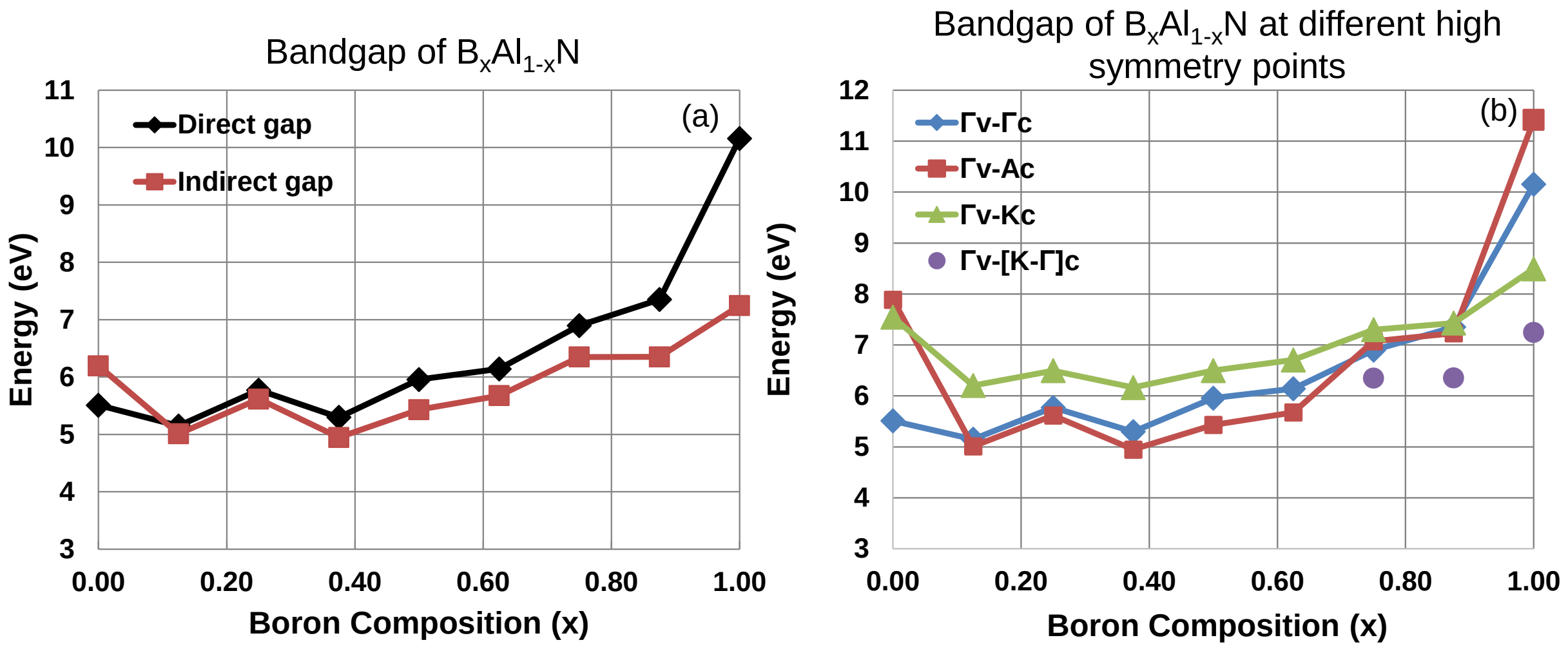

Figure 5 (a) The direct and indirect bandgaps of BAlN as functions of boron composition. (b) The bandgaps of BAlN for different transitions between chosen symmetry points.

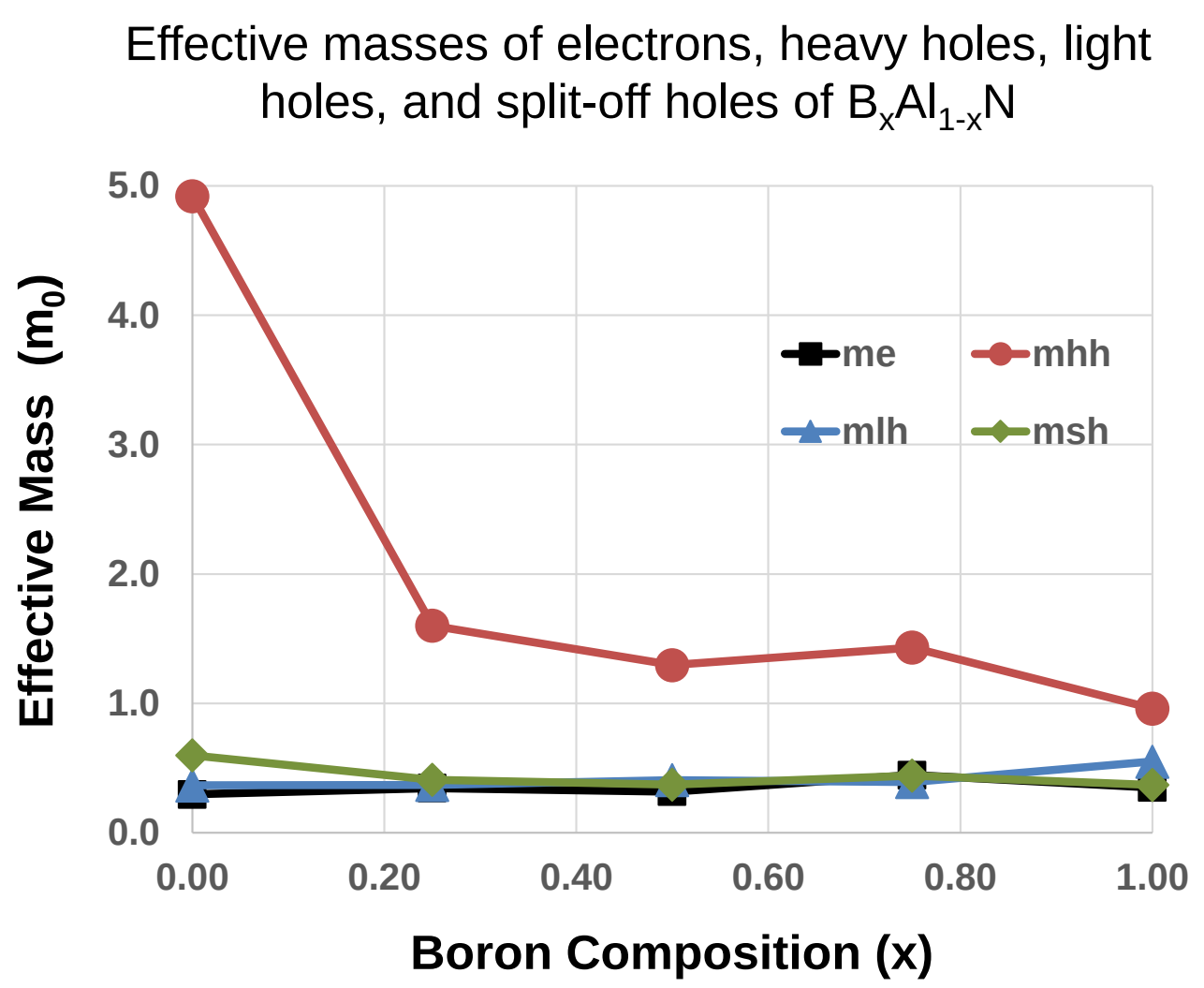

Figure 6 Effective masses of BAlN with different boron compositions.

## Conclusion

In this work we report the electronic band structures of BAlN and BGaN alloys using DFT. We also report some band structure dependent properties such as the bandgap and the direct-to-indirect bandgap transition and electron and hole effective masses. We find that the direct to indirect transition for BGaN and BAlN are near boron compositions of 44% and 24%, respectively. A huge drop in hh effective mass is observed for both BGaN and BAlN at boron compositions below 25%. A similar –but smaller– drop for lh is found in BGaN only. A figure showing the bangaps of nitrides in the wurtzite phase and their lattice parameters is made summarizing the results. This work should guide material growth experiment as interest in boron-containing nitrides increases.